\documentclass[sigconf]{acmart}

\AtBeginDocument{%
  }

\setcopyright{acmcopyright}
\copyrightyear{2018}
\acmYear{2018}
\acmDOI{XXXXXXX.XXXXXXX}

\acmConference[Conference acronym 'XX]{Make sure to enter the correct
  conference title from your rights confirmation emai}{June 03--05,
  2018}{Woodstock, NY}
\acmPrice{15.00}
\acmISBN{978-1-4503-XXXX-X/18/06}

\usepackage{subcaption}
\usepackage{url}

\begin{document}

\title{LOUC: Leave-One-Out-Calibration Measure for Analyzing Human Matcher Performance}

\author{Matan Solomon}
\email{matan-so@campus.technion.ac.il}
\affiliation{%
  \institution{Technion -- Israel Institute of Technology}
  \city{Haifa}
  \country{Israel}
}

\author{Bar Genossar}
\email{sbargen@campus.technion.ac.il}
\affiliation{%
	\institution{Technion -- Israel Institute of Technology}
	\city{Haifa}
	\country{Israel}}

\author{Roee Shraga}
\email{r.shraga@northeastern.edu}
\affiliation{%
	\institution{Northeastern University}
	\city{Boston}
	\state{MA}
	\country{USA}}

\author{Avigdor Gal}
\email{avigal@technion.ac.il}
\affiliation{%
  \institution{Technion -- Israel Institute of Technology}
  \city{Haifa}
  \country{Israel}}

\renewcommand{\shortauthors}{}

\newcommand{\ms}[1]{\textcolor{red}{MS: {#1}}} 
\newcommand{\bg}[1]{\textcolor{teal}{BG: {#1}}} 
\newcommand{\rs}[1]{\textcolor{blue}{RS: {#1}}} 
\newcommand{\ag}[1]{\textcolor{green}{AG: {#1}}} 

\begin{abstract}
	Schema matching is a core data integration task, focusing on identifying correspondences among attributes of multiple schemata. Numerous algorithmic approaches were suggested for schema matching over the years, aiming at solving the task with as little human involvement as possible. Yet, humans are still required in the loop -- to validate algorithms and to produce ground truth data for algorithms to be trained against. In recent years, a new research direction investigates the capabilities and behavior of humans while performing matching tasks. Previous works utilized this knowledge to predict, and even improve, the performance of human matchers. In this work, we continue this line of research by suggesting a novel measure to evaluate the performance of human matchers, based on calibration, a common meta-cognition measure. The proposed measure enables detailed analysis of various factors of the behavior of human matchers and their relation to human performance. Such analysis can be further utilized to develop heuristics and methods to better asses and improve the annotation quality.
\end{abstract}

\maketitle

\section{Introduction}

Schema Matching (SM), a central data integration task, focuses on finding correspondences between schema attributes. SM serves as a fundamental data preparation step, used, for example, for merging multiple data sources~\cite{Bernstein2011}. The problem has been broadly researched, resulting in multiple SM solutions with varying degrees of human involvement, ranging from completely manual~\cite{Crowdmap,zhang2018reducing} to fully algorithmic approaches~\cite{koutras2021valentine,shraga2020adnev}. 
A significant amount of work also aimed at developing robust semi-automated approaches, where a matching algorithm generates correspondences that are then validated by one or more human experts~\cite{Falconer2007,fan2014hybrid,McCann2008}. All in all, human-assisted data integration requires both efficient algorithmic matchers and a well grounded understanding of human matching, with the latter being the focus of this work.

The need for human-in-the-loop schema matching does not end with validating the generated results. Powerful matching algorithms that rely on supervised machine learning require large amounts of training data. For this reason, human involvement is also necessary to generate reliable ground truth data for training the algorithms. Data annotation can be a costly process, especially when performed by experts~\cite{li2017human,li2019user}. Hence, relying solely on matching experts for this task can be challenging, leading to an increasing interest in the utilization of crowd-sourcing platforms in recent years.

Relying on non-expert human annotators challenges the assumption that humans are generally better matchers than algorithms~\cite{shraga2021learning}. Crowdsourcing platforms (e.g., Prolific Academic~\cite{ProlificAcademic} and CrowdFlower~\cite{CrowdFlower}) serve as a good quick source for producing big amounts of labeled data. Yet, the annotations are usually of low quality and conducted without much attention~\cite{Ross2010,ghazarian2010automatic,callison2009fast}. These disadvantages led to the realization that to 
optimize crowdsourcing annotations, further knowledge about the way humans match is needed. In recent years, a new research direction focuses on the capabilities and behaviors of humans in the matching process, and the various ways to utilize this knowledge to predict and improve their performance~\cite{shraga2020mind}. Previous works in this research area use machine learning models to characterize and understand human matchers~\cite{shraga2021learning}, focusing on predicting the capabilities of human annotators~\cite{shraga2022humanal} and to filter their choices~\cite{shraga2022powarematch}, resulting in a significant improvement in the quality of the labeled data.

Analyzing the behavior of human matchers, in contrast to algorithmic matchers, necessitates an understanding of the cognitive mechanisms involved in executing the matching task. The interdisciplinary nature of the problem creates fertile ground for collaborations between computer and behavioral sciences. Previous works borrowed well-established models from behavioral sciences in general and meta-cognition in particular, to map the psychological biases of human matchers~\cite{ackerman2019cognitive}. Taking into account the psychological aspects of matching opens up an entirely new dimension also in the evaluation of human-based matching. Traditional evaluation measures such as accuracy and precision~\cite{bellahsene2011evaluating} may not fully capture the complexity of the behavior of human matchers. Therefore, incorporating measures such as response time and reported confidence level can offer a more comprehensive evaluation of matching performance, a crucial step towards improving it.

We believe that developing new measures to assess human input is crucial for improving the efficiency and reliability of human-assisted data integration. Therefore, in this work we focus on {\em calibration}, a well-known meta-cognition evaluation measure~\cite{hattie2013calibration} that assesses the discrepancy between our judgment and the accuracy of the match. Calibration is used to evaluate machine learning models (which were trained using human annotations) ability to generate probability estimates representative of the true correctness likelihood~\cite{guo2017calibration}. We aim to adapt and analyze the notation of calibration to enhance the understanding of human input in the specific task of schema matching. Specifically, we propose {\em LOUC (leave one out calibration)}, a measure that enables a more detailed analysis and provides insights about the different factors that impact the human performance. We present an initial empirical evaluation of the measure using human-annotated data.


The rest of the paper is organized as follows. In Section~\ref{sec: SM model} we provide a model of schema matching, and present the data for our empirical evaluation. Then, in Section~\ref{sec: calibration based measures} we present the calibration measure, and our suggested measure, both escorted by empirical results over our data. Section~\ref{sec: related work} reviews previous works in the field, and finally, we discuss the limitations and future work in Section~\ref{sec: Limitations and Challenges}. All of the data used in our experiments, as well as the code to reproduce the results, are available in our Git repository\footnote{\url{https://github.com/MatanSolomon01/LOUC-LeaveOneOutCalibration}}.

\section{Human matching model}
\label{sec: SM model}
In this section we introduce a general schema matching model (Section~\ref{sec:sm}), and discuss human matching (Section~\ref{sec:hm}), traditional evaluation measures (Section~\ref{subsec: Traditional measures}), and offer a detailed description of experiments we ran with human matchers (Section~\ref{subsec: HumanAnnotators}).
\subsection{Schema Matching Model}\label{sec:sm}
Schema matching involves finding correspondences between two schemata, $S, S^{\prime}$, by aligning their attributes, denoted as $ \{a_1, \dots, a_m\} $ and $\{b_1, \dots, b_n\} $, respectively. The result of the matching process can be formalized as a \textit{similarity matrix $ M(S, S') \in [0,1]^{m\times n} $} (denoted as $ M $ when context is clear), where the rows and columns represent the attributes of the schemata, and each entry $ M_{i,j} \in [0,1] $ represent the similarity between $ a_i \in S $ and $ b_j \in S' $.

Schema matching can be positioned as a binary classification problem, assigning a binary label for each pair of attributes $(a_i, b_j)$. We use $ \sigma $ to denote a \textit{match}, a subset of $ M $'s entries, obtained from $ M $ according to some rule, and represent the set of attribute pairs we binarly consider as a match. For example, a match $\sigma_M = \{(i,j) | M_{i,j} > \delta\}$ can be obtained using a threshold $ \delta $ over the entries.


To evaluate the performance of a match, we rely on a matrix $M^e \in \{0,1\}^{m\times n}$, which serves as a reference point. The matrix is constructed such that $M^e_{i,j}=1$ if the pair $(a_i, b_i)$ is part of the reference match, and $M^e_{i,j}=0$ otherwise. Reference matrices are constructed and refined by matching experts over years, and thus are usually considered as ground truth.


\subsection{Human Matching}
\label{sec:hm}
Crowdsourcing platforms are being increasingly used for data annotation due to the need for large amounts of annotated data and the accessibility of the internet. However, participants in crowdsourcing tasks are not typically domain experts~\cite{zhang2018reducing,li2019user} and may have cognitive biases~\cite{ackerman2019cognitive}, which may lead to low-quality labels~\cite{Ross2010,ghazarian2010automatic,callison2009fast}. 


{\em InCognitoMatch (ICM)}~\cite{shraga2020incognitomatch} is a matching interface, used in our experiments (see Section~\ref{subsec: HumanAnnotators}). {\em ICM} is a cognitive-aware crowdsourcing application tailored for matching tasks. 
This platform allows for the investigation of cognitive biases, including over-confidence and conformism, among others. For illustration and additional details, see The {\em ICM} Web site~\cite{inCognitoMatch}. 
Similar interfaces, {\em e.g.}, Ontobuilder~\cite{ontobuilder}, 
investigate various aspects of the matching task~\cite{shraga2022powarematch,shraga2021learning, zhang2018reducing}. 

When deciding whether two attributes correspond, both human annotators and algorithms can consider various attribute characteristics. Attribute names are often a good elementary indication as to whether the pair correspond. Attribute data types ({\em e.g.}, integer, string) can also be useful to distinguish between attributes, as well as the location of each attribute in its schema hierarchy. Furthermore, table content can be compared to determine similarity. 

As part of the annotation process, the interface gathers participant responses. Resembling the similarity scores provided by matching algorithms, human annotators provide confidence scores, indicating how confident they are about whether a pair corresponds. The reported confidence is on a scale of $[0, 100]$, where 0 is a complete confidence in a non-match, 100 is a complete confidence in a match, and 50 indicates neutrality. In our experiments, we normalize the confidence scores to $[0,1]$. Other aspects of human matching, {\em e.g.}, response time (seconds), were used in previous works to improve the annotations by filtering out wrong decisions, and even to predict certain types of matching expertise~\cite{shraga2021learning, shraga2022humanal, shraga2022powarematch}. In the scope of this work, we focus on the reported confidence and response time.

Consider a set of human annotators $H = \{h_1, h_2, \cdots, h_K\}$, which are presented with an (ordered) series of matching questions $Q = (q_1, q_2, \cdots, q_B)$. 
The answer of human annotator $h$ to matching question $q = (a_i, b_j)$ is denoted by $A^h_q$, and is constructed of reported confidence and response time $A^h_q = (C^h_q, T^h_q)$. A similarity matrix $M^h$ can be obtained for each matcher $h$ by assigning $M^h_{i,j} = C^h_q$ such that $q=(a_i, b_j)$. Note that such a similarity matrix is  sparse, as the annotator is typically not presented with all possible pairs. A match $\sigma^h$ can be constructed by $\sigma^h = \{(i,j)|\exists q=(a_i, b_j)\in Q, C^h_q > \delta = 0.5\}$.
Recall that the reported confidence is in the range of $[0,1]$, where values above (below) 0.5 indicate a match (non-match). Following~\cite{shraga2022humanal}, to separate the binary decision from its associated confidence, we convert each reported confidence to 
its associated {\em normalized confidence} as follows.
\begin{equation}\label{equ: confidence normalization}
	\tilde{C}_q^h = 2\cdot\mid C_q^h - 0.5\mid
\end{equation}

\subsection{Traditional Evaluation Measures}
\label{subsec: Traditional measures}
Before diving into calibration-based measures, we present traditional matching evaluation measures. 
As a binary classification task (see Section~\ref{sec:sm}), we define the basic measures of accuracy and precision,\footnote{Recall is not well defined when annotators are given a set of pre-defined questions, so we focus on precision and accuracy only.} aggregative measures that summarize the performance of a human matcher. For the scope of this work accuracy and precision are defined with respect to a subset of questions $Q^\prime\subseteq Q$. Accuracy (Eq.~\ref{equ: Accuracy}) is the proportion of correctly labeled questions, both positive and negative, and precision (Eq.~\ref{equ: Precision}) looks only at the positives.



\begin{equation}\label{equ: Accuracy}
	Acc^h(Q^\prime) = \frac{|\{q|q\in Q^\prime, q=(a_i, b_i), \mathbb{I}_{(i,j)\in\sigma^h} = \mathbb{I}_{(i,j)\in\sigma_{M^e}}\}|}{|Q^\prime|}
\end{equation}
\begin{equation}\label{equ: Precision}
	P^h(Q^\prime) = \frac{|\sigma^h \cap M^{e^+}\cap Q^\prime|}{|\sigma^h \cap Q^\prime|}
\end{equation}
\noindent where $\mathbb{I}(\cdot)$ denotes an indicator function and $M^{e^+}$ is the set of positive ($1$) entries of $M^e$. 

%
%

We also adopt confusion matrices notation, denoting by {\em True Positive ($TP$)} and {\em False Positive ($FP$)} the set of matching decision for which the annotator's binary label was positive ({\em i.e.}, a match), and she was correct, or mistaken, respectively. We also use {\em True Negative ($TN$)} to denote the set of decisions that the annotator correctly labeled as non-matches, and {\em False Negative ($FN$)} to denote the set of questions that the annotator mistakenly labeled as non-matches.

Binary measures evaluate algorithmic matchers and are also useful for providing a good assessment of the human matcher's performance. Human matchers, however, are affected by cognitive biases that algorithms avoid. Thus, following the meta-cognition literature~\cite{ackerman2016metacognition,hattie2013calibration,ackerman2019cognitive}, other measures should be added to allow a better understanding of human matcher behavior. Utilizing this cognitive behavioral knowledge to evaluate performance is the first step to better methods for improving the human labeling quality.

\subsection{Experimenting with Human annotators}
\label{subsec: HumanAnnotators}
We used the ICM interface (see Section~\ref{sec:hm}) to annotate a pair of schemata. We used the publicly available Purchase Order dataset~\cite{do2002coma}, which comprises schemata with a substantial amount of information content, including data types and instance examples.

The experiment was conducted twice. In one experiment we used the crowdsourcing platform of Prolific Academic~\cite{ProlificAcademic} to direct participants to ICM. Our inclusion criteria involved selecting only desktop users with English as their primary language and at least a 95\% approval rate.
\footnote{Approval rate is the crowdsourcing tasks rate for which the participant was approved.} In the second experiment, participants were students in a databases management course at the \emph{Technion} 
(Haifa, Israel). The students primary language is either Hebrew or Arabic, and during the experiment they were in their second year of bachelor's degree in data science or information systems. Each participant in both experiments completed the same series of 30 matching decisions, for which we collected the decision time and confidence. Personal information of participants, such as age and gender, was not utilized or shared due to privacy constraints. After filtering participants with low performance and experiments with technical issues or errors, we gathered 147 valid participants, resulting in a dataset of 4410 matching decisions.


The participants spent an average of 7.6 minutes in the matching task, where each matching question was answered within 15 seconds on average. Of the 30 attribute pairs presented to participants, half were matches, and half were non-matches. Yet, the participants tended to label the pairs as matches, and the dataset includes 2781 (63\%) positive (match) predicted labels and only 1629 (37\%) negative (non-match) predicted labels. Ignoring the binary decision, and looking only at the normalized reported confidence (see Eq.~\ref{equ: confidence normalization}), the average confidence score reported by the participants is 0.508. 

\section{Calibration Based Measures}
\label{sec: calibration based measures}
We next offer an in-depth analysis, both theoretical and empirical, of human matcher performance using the calibration measure (Section~\ref{subsec: The calibratin measure}), and introduce a new measure, {\em LOUC}, offering interesting new insights on human matcher behavior (Section~\ref{sec: LOUC}).
\begin{figure*}[htpb]
	\captionsetup[subfigure]{justification=centering}
	\begin{minipage}[b]{0.245\textwidth}
		\includegraphics[width=\linewidth]{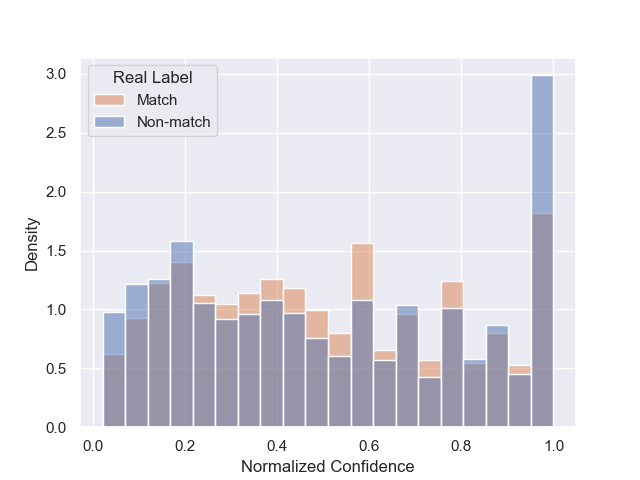}
		\subcaption{Confidence by real label\\$ $}
		\label{fig: confidence-distribution-by-real-label}
	\end{minipage}
	\begin{minipage}[b]{0.245\textwidth}
		\includegraphics[width=\linewidth]{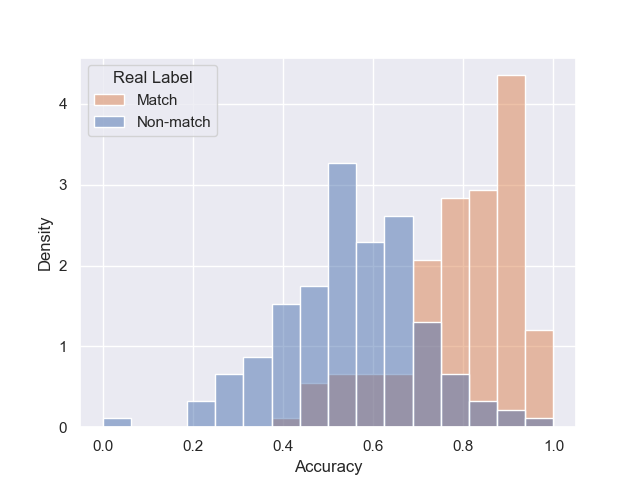}
		\subcaption{Accuracy by real label\\$ $}
		\label{fig: accuracy-distribution-by-real-label}
	\end{minipage}
	\begin{minipage}[b]{0.245\textwidth}
		\includegraphics[width=\linewidth]{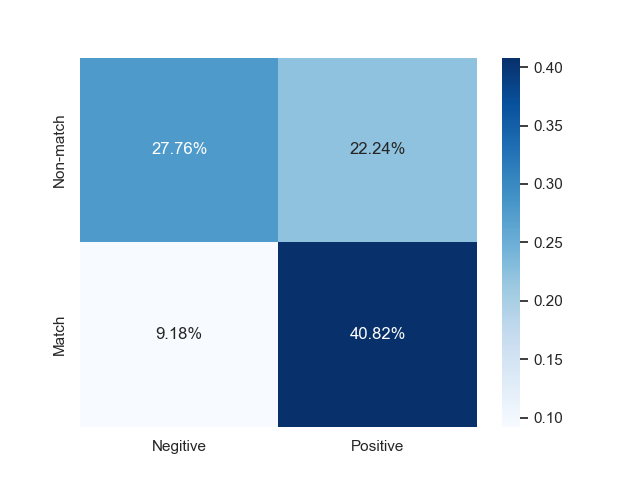}
		\subcaption{Confusion matrix\\$ $}
		\label{fig: confusion matrix}
	\end{minipage}
	\begin{minipage}[b]{0.245\textwidth}
		\includegraphics[width=\linewidth]{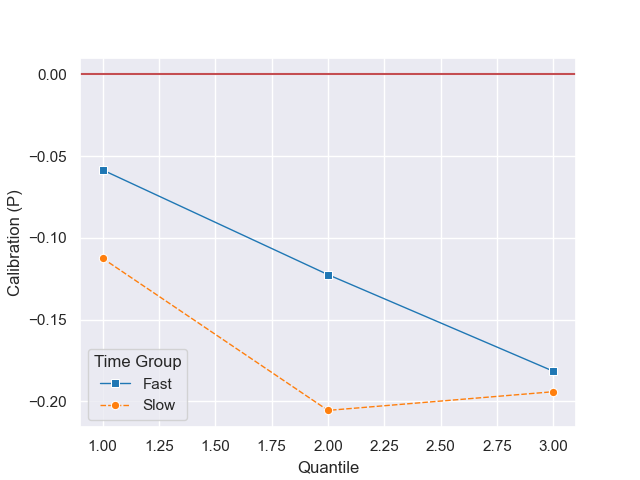}
		\subcaption[Calibration (P) by response time quantile, per time group]{Calibration (P) by response time quantile, per time group}
		\label{fig:calibration-p-by-answering-time-quantile-per-time-group}
	\end{minipage}
	\centering
	\caption{Calibration Analysis}
\end{figure*}

\subsection{The Calibration Measure}
\label{subsec: The calibratin measure}
In our settings (Section \ref{subsec: HumanAnnotators}) human matchers provide confidence scores for matching pairs. 
We view the confidence score as a measure of match decision correctness. 
However, in real-life scenarios, a reference match does not always exist, and we must identify matchers for whom we can rely on their reported confidence. 
In this context, trustworthy matchers are matchers that report confidence levels that closely align with their actual performance. Such matchers are termed {\em  calibrated}~\cite{shraga2021learning} and they are valuable in the absence of ground truth, allowing us to correctly assess their performance by reported confidence. 

Uncalibrated matchers can be categorized into two groups, under-confident and over-confident. The former includes matchers that tend to report confidence levels that are lower than their actual performance, underestimating their performance. 
The latter includes matchers that provide confidence levels that are higher than the actual performance, overestimating their performance. Identifying the group of a matcher can help improve labeling quality by adjusting the reported confidence levels. 

The calibration measure
is a well known metric in the field of meta-cognition~\cite{ackerman2016metacognition,hattie2013calibration}. The measure quantifies the success of a matcher in self-assessing performance, enabling us to classify 
matchers as under-confident, calibrated, or over-confident. Calibration is an aggregative measure, and it is essentially the difference between the mean of the reported normalized confidence and the actual performance, by some traditional measure (see Section~\ref{subsec: Traditional measures}). For example, an accuracy-based calibration for a matcher $h$, calculated over a subset of questions $Q^\prime \subseteq Q$ is computed as follows. 
\begin{equation}\label{equ: cal (acc)}
	Cal^h_{Acc}(Q^\prime) = \tildebar{C}^h(Q^\prime) - Acc^h(Q^\prime) \qquad \tildebar{C}^h(Q^\prime) = \frac{1}{|Q^\prime|}\sum_{q\in Q^\prime} \tilde{C}^h_{q}
\end{equation}

The mean reported normalized confidence should be calculated only over decisions that are relevant to the performance measure. For example, the precision ignores $TN$ and $FN$. 
We use $\tildebar{C}(Q^\prime)$ to denote the mean normalized confidence over a subset of question  $Q^\prime \subseteq Q$, and indicate the group of decisions for which we compute the mean using a subscript. For example, $\tildebar{C}_{TP, FP}(Q^\prime)$ denotes the mean normalized confidence over $TP$ or $FP$ decisions in $Q^\prime$.

Eq.~\ref{equ: cal (P)} presents calibration over precision. The simplicity and informativeness of the measure make it an effective tool for evaluating human matchers performance.
\begin{gather}\label{equ: cal (P)}
	Cal^h_{P}(Q^\prime) = \tildebar{C}_{TP, FP}^h(Q^\prime) - P^h(Q^\prime)\\
	\tildebar{C}_{TP, FP}^h(Q^\prime) = \frac{1}{|(TP\cup FP)\cap Q^\prime|}\sum_{q\in (TP\cup FP)\cap Q^\prime} \tilde{C}^h_{q}\nonumber
\end{gather}



\begin{figure}[tpb]
	\captionsetup[subfigure]{justification=centering}
	\begin{subfigure}[b]{0.23\textwidth}
	\includegraphics[width=\linewidth]{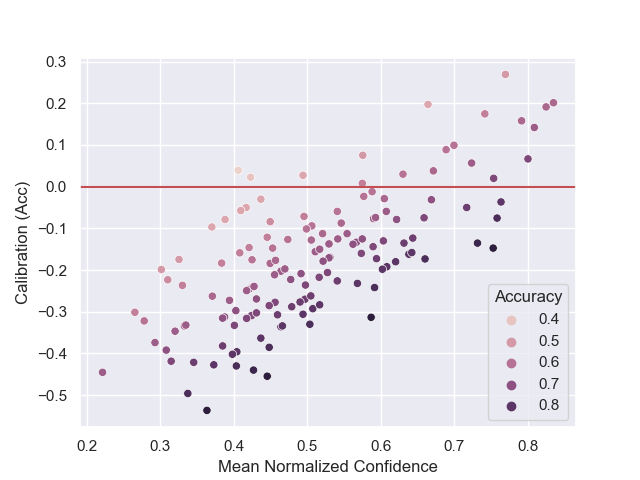}
	\caption[Calibration (Acc) by mean normalized confidence and accuracy]{Calibration (Acc) by mean confidence and accuracy.}
	\label{fig:calibration-acc-by-mean-normalized-confidence}
	\end{subfigure}
	\hfill
	\begin{subfigure}[b]{0.23\textwidth}
	\includegraphics[width=\linewidth]{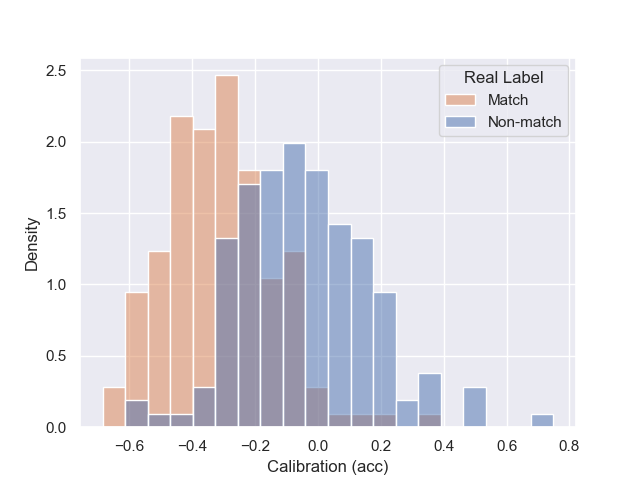}
	\caption[Calibration (Acc) distribution by real label]{Calibration (Acc) by real label.}
	\label{fig:calibration-acc-distribution-by-real-label}
	\end{subfigure}
	\caption{Calibration by confidence and by the real label.}
	\vskip-.15in
\end{figure}

\noindent\textbf{Calibration Empirical Analysis:} To provide an initial understanding of calibration behavior, consider Figure~\ref{fig:calibration-acc-by-mean-normalized-confidence}, which presents the calibration (acc) by the normalized mean confidence and accuracy, for all participants and questions. 
The points below the red $y=0$ line, represent annotators whose mean confident level is below their accuracy. These are the under-confident annotators. The points above the line represent over-confident annotators. A point near the red line represents a participant that is more calibrated. For a given mean confident level, we can see that the calibration decreases towards under-confidence as accuracy increases. The reason is that for annotators with a fixed confidence level, a more accurate annotator underestimates performance more. The overall trend indicates that more confident annotators tend to be characterized with positive and increasing calibration. The reason is that a more confident annotator is more likely to overestimate performance.

Figure~\ref{fig:calibration-acc-distribution-by-real-label} displays the calibration distribution of annotators, separated into true matches and non-matches. 
We can see that for non-match decisions, the annotators are more calibrated (with values around $0$) than for match decisions. This observation is not trivial, since the annotators do not know the true label while matching. We offer two possible explanations for this phenomenon: 1) the confidence levels are lower for true matches, or 2) the annotators are more accurate when answering true match questions.

\begin{figure*}[tpbh]
	\captionsetup[subfigure]{justification=centering}
	\begin{minipage}[b]{0.3\textwidth}
		\includegraphics[width=\linewidth]{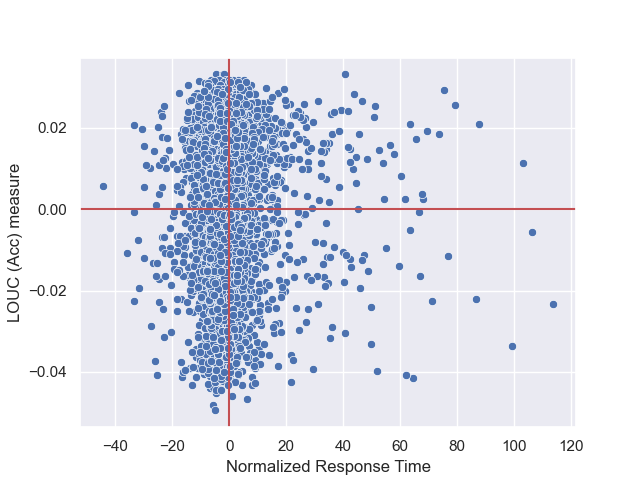}
		\subcaption{LOUC (Acc) by \\Response Time}
		\label{fig: LOUC (Acc) Measure by Normalized Response Time}
	\end{minipage}
	\begin{minipage}[b]{0.3\textwidth}
		\includegraphics[width=\linewidth]{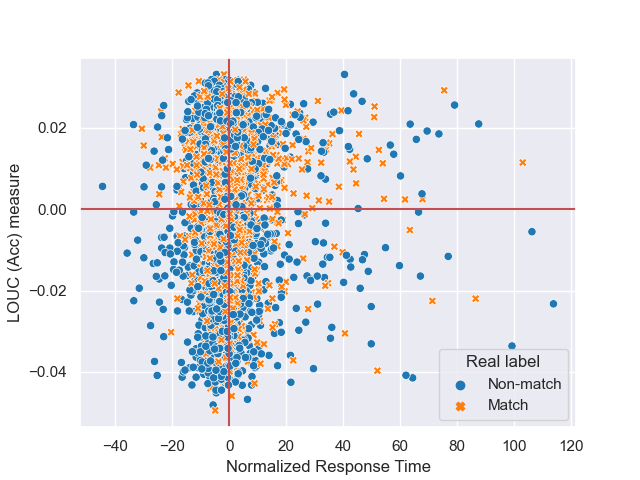}
		\subcaption{LOUC (Acc) by Response Time \\and Real Label}
		\label{fig: LOUC (Acc) Measure by Normalized Response Time and Real Label}
	\end{minipage}
	\begin{minipage}[b]{0.3\textwidth}
		\includegraphics[width=\linewidth]{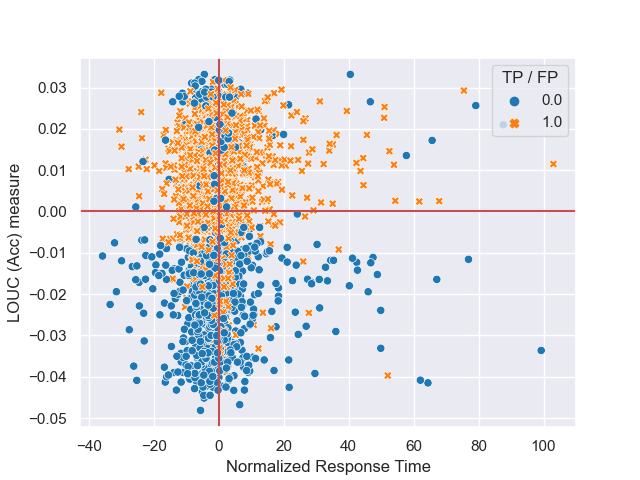}
		\subcaption{LOUC (Acc) by Response Time and Positive Predicted Label}
		\label{fig: LOUC (Acc) Measure by Normalized Response Time and Positive Predicted Label}
	\end{minipage}
	\begin{minipage}[b]{0.3\textwidth}
		\includegraphics[width=\linewidth]{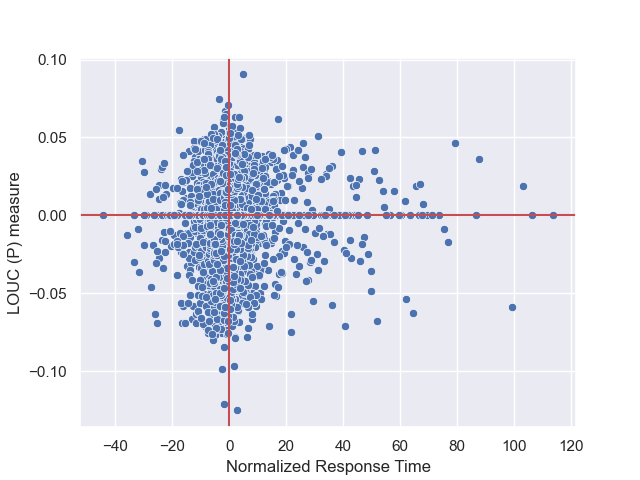}
		\subcaption{LOUC (P) by \\Response Time}
		\label{fig: LOUC (P) Measure by Normalized Response Time}
	\end{minipage}
	\begin{minipage}[b]{0.3\textwidth}
		\includegraphics[width=\linewidth]{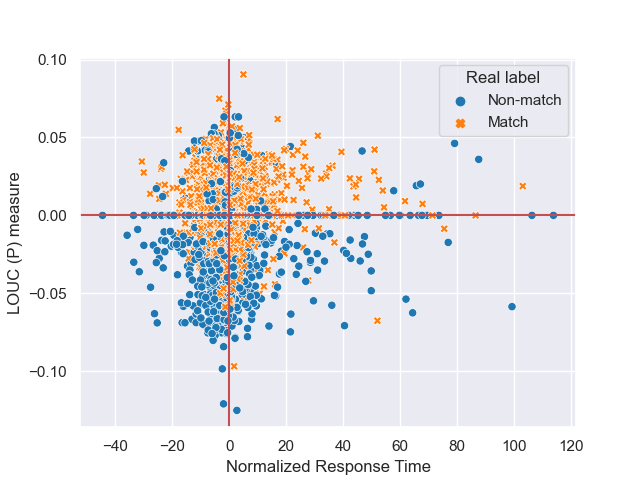}
		\subcaption{LOUC (P) by \\Response Time and Real Label}
		\label{fig: LOUC (P) Measure by Normalized Response Time and Real Label}
	\end{minipage}
	\begin{minipage}[b]{0.3\textwidth}
		\includegraphics[width=\linewidth]{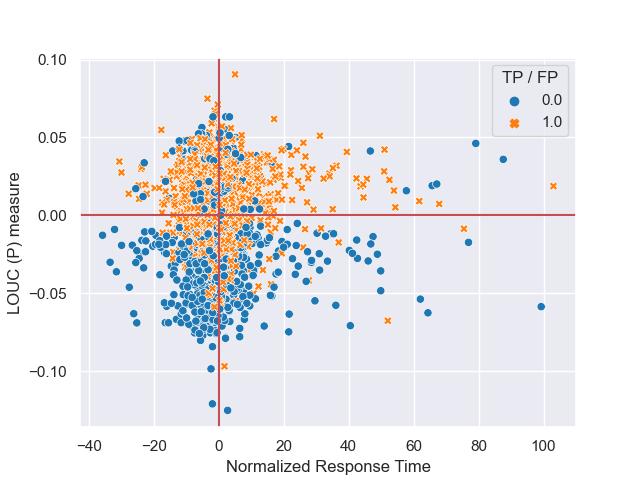}
		\subcaption{LOUC (P) by  Response Time and Positive Predicted Label}
		\label{fig: LOUC (P) Measure by Normalized Response Time and Positive Predicted Label}
	\end{minipage}
	\caption{Accuracy-based LOUC (Figures \ref{fig: LOUC (Acc) Measure by Normalized Response Time}, \ref{fig: LOUC (Acc) Measure by Normalized Response Time and Real Label} and \ref{fig: LOUC (Acc) Measure by Normalized Response Time and Positive Predicted Label}) and Precision-based LOUC (Figures \ref{fig: LOUC (P) Measure by Normalized Response Time}, \ref{fig: LOUC (P) Measure by Normalized Response Time and Real Label} and \ref{fig: LOUC (P) Measure by Normalized Response Time and Positive Predicted Label})}
	\vskip-.1in
\end{figure*}


Figure~\ref{fig: confidence-distribution-by-real-label} presents the normalized confidence distributions by the real label. The distributions are similar, which is a notable finding, indicating that while calibration was beneficial for distinguishing between match and non-match questions, confidence alone does not seem to be a reliable indicator.
Figures \ref{fig: accuracy-distribution-by-real-label} and \ref{fig: confusion matrix} present the distribution of the accuracy by the real label. Figure \ref{fig: accuracy-distribution-by-real-label} shows that the annotators are more accurate when answering true match questions. Figure \ref{fig: confusion matrix}, which presents how the decisions are distributed across the $TP$, $TN$, $FP$, $FN$ groups reveals why -- the annotators tend to annotate questions as matches, which is reflected by the high rates in the "Positive" column and causes them to be more accurate when the true label is indeed a match. The calibration measure captures this phenomenon, which serves as evidence of its predictive power in revealing the true label.


Next, 
we analyze the relationship between response time and
calibration. 
We divide the annotators into 2 groups of ``slow'' and ``fast'' matchers, according to their mean response time. The slow annotators are 35 seconds slower than the average, whereas the fast annotators are 0.8 faster than the average. The bias towards slow participants may be a result of the (implicit) lower bound of response time. 
Response time is normalized by question, considering the mean response time of a question across annotators and the mean response time of the annotator. The specific normalization is given in Section~\ref{sec: LOUC}. We further discretize the calibration level by three quantiles of normalized response time, such that each quantile consists of 10 matching decisions, aiming to reflect different difficulty levels. We calculate the annotator's calibration (precision based) for each group. Finally, we average calibration levels separately for the fast and slow groups of participants.


The results are presented in Figure~\ref{fig:calibration-p-by-answering-time-quantile-per-time-group}. On average, participants are under-confident as both lines lie below the $y=0$ line. The line of fast-annotators is above that of the slow annotators, implying that faster annotators tend to be more calibrated (\emph{i.e.}, less under-confident) than slower annotators. Lastly, we observe a decreasing trend in calibration as response time increases, suggesting that participants are less calibrated for possibly difficult questions that take longer to answer. This suggests longer response times make it harder to assess self-performance, tying it with calibration.


\subsection{LOUC: Leave-One-Out Calibration}
\label{sec: LOUC}
%
Calibration is an aggregative measure, which is applied over a series of decisions. As the set of decisions becomes smaller, calibration becomes more sensitive to  outliers. 
In particular, it is impossible to test the calibration for a single decision, which can be very useful, for example, for isolating the characteristics that affect the calibration in the finest granularity level of a single question. Aiming to fill this gap we now present {\em LOUC -- Leave-One-Out} calibration measure. The LOUC measure enables the evaluation of the effect of each matching decision over the calibration level of an annotator. It essentially allows testing the influence of characteristics in a granularity level of a single question over the calibration, without compromising the stability and reliability of the measure. The fine analysis enabled by this measure serves as the basis for a variety of future works that focus on improving annotators' performance. For example, knowing matching questions characterization for which the annotators are more calibrated, we could construct annotation tasks that direct the annotators into more calibrated behavior.

\begin{sloppypar}
We start by normalizing matchers response time to create a common baseline for all matchers and avoid inherent temporal bias among them. The average response time per question, which serves as an indication of its difficulty level, is given by $\bar{T}_q = \frac{1}{K} \sum_{i=1}^{K} T_q^{h_i}$.
\end{sloppypar}

Then, for each human matcher, we define its average answering delta, which quantifies a matcher's temporal performance with respect to the average. The average answering delta (Eq.~\ref{equ: average asnwering delta}, left) is then used to define the response time normalization (Eq.~\ref{equ: average asnwering delta}, right).
\begin{equation}\label{equ: average asnwering delta}
	\Delta^h = \frac{1}{B} \sum_{i=1}^{B} T^h_{q_i} - \bar{T}_{q_i}, \quad \tilde{T}^h_q = (T_q^h - \bar{T}_q) - \Delta^h
\end{equation}



The normalized response time is essentially the matcher’s response time deviation in the current question, with respect to his average deviation. For example, consider a matcher that is faster than the average matcher by 5 seconds ($\Delta^h=-5$). Then, in some question $q$, that matcher was 7 seconds faster than the average ($T_q^h-\bar{T}_q=-7$). This means that in this question the matcher was 2 seconds faster than the self-average ($\tilde{T}_q^h=-7-(-5)=-2$). Using this normalization, we can compare the effect of response times while neutralizing the differences between annotators.

The LOUC measure quantifies the extent to which a matching decision contributes to matcher's calibration. 
 It is the difference, in absolute values, between the calibration level calculated over all the questions, to the calibration level calculated over the questions other than $q$.
We use $\varphi^h(q)$ to denote the LOUC of a decision made by annotator $h$ for question $q$, and we indicate the traditional measure in use, using a subscript. For example $\varphi^h_{Acc}(q)$ is the accuracy-based LOUC value of $h$'s decision regarding question $q$. LOUC, with respect to a measure $E$ (accuracy/precision) is given by:
\begin{equation}\label{equ: LOUC}
	\varphi^h_{E}(q) = |Cal^h_{E}(Q)| - |Cal^h_{E}(Q\setminus\{q\})|
\end{equation}
\noindent For example, precision-based LOUC uses $Cal^h_{P}(Q)$ (Eq.~\ref{equ: cal (P)}).


Recall that a calibrated matcher is a matcher whose calibration level is close to 0. To this point, LOUC uses the calibration with absolute values, which disregards the sign of the calibration, and essentially evaluates the extent to which an annotator is calibrated, without the notion of under/over confident. In this context, we provide the following interpretation for LOUC. Negative LOUC decisions decrease the absolute value of the calibration and thus make the annotator more calibrated. Positive LOUC decisions increase the absolute value of the calibration. All in all, calibration-wise ``good'' questions are those with negative LOUC values. 

It is worth noting that the value of each decision is determined by how much it helps to reduce the calibration level of other decisions towards zero. As all the other decisions are already included in this calculation, this produces a stable measure unaffected by outliers, and thus we anticipate that the resulting values will be small. Furthermore, by measuring the contribution of each decision, rather than just the calibration baseline, we obtain a normalized measure that accounts for variations between annotators. This enables focusing solely on the effect of question characteristics and ignore any discrepancies between different matchers.   

Lastly, note that precision based LOUC (Eq. \ref{equ: LOUC}) inherently assign a value of 0  to some decisions. Decisions that are not used to calculate the calibration do not contribute to its value. For example, a $TN$ decision $q\in TN$ would have a $ \varphi^h_P(q) $ of 0. That is because the left-term (precision-based) calibration does not consider TN questions, and the right-term calibration ignores it anyway. Therefore, the calibration levels are equal, and the contribution is $\varphi^h_P(q)=0$. 

\vspace{.05in}\noindent\textbf{LOUC Empirical Analysis:} We proceed by providing empirical results of the measure over our data and discuss several interesting phenomena it presents. Figures~\ref{fig: LOUC (Acc) Measure by Normalized Response Time} and~\ref{fig: LOUC (P) Measure by Normalized Response Time} present the accuracy and precision based LOUC values of all decisions, by the normalized response time. The overall trend (which is more significant in the precision-based LOUC) suggests that as further the normalized response time is from 0, as closer to 0 the LOUC value gets. Close to 0 LOUC decisions does not contribute to the annotator calibration, and in these decisions, the annotator demonstrates a typical (but not necessarily good) calibration ability. The results suggest that as the response time becomes less typical for the matcher (normalized response time far from 0), we should expect a more typical behavior. On the other hand, variations in the calibration ability of the annotator are more likely to be revealed when the response time is typical (normalized response time close to 0).

For ease of presentation, the remaining figures use the same base graph (LOUC values by normalized response time) while demonstrating different question characteristics. Figures~\ref{fig: LOUC (Acc) Measure by Normalized Response Time and Real Label} and ~\ref{fig: LOUC (P) Measure by Normalized Response Time and Real Label} show the distribution of the decisions by the real label (true match/non-match), suggesting that (true) non-match decisions are more concentrated in the negative LOUC values, whereas (true) match decisions are more concentrated in the positive LOUC values.\footnote{This result is more significant in the precision-based LOUC, but also appears in the accuracy-based, as for non-match decisions, the majority (55\%) have negative LOUC values, and for match decisions, the majority (62\%) have positive LOUC values.} This is an interesting observation, recalling that 
in Figure~\ref{fig:calibration-acc-distribution-by-real-label} annotators were shown to be less calibrated for true match decisions. Here, we see that true match decisions have positive LOUC values, moving the calibration levels away from 0, in contrast to true non-match decisions, which contribute to making the annotators more calibrated. This indicate that the LOUC measure preserves the behavior of the calibration, even at a finer granularity level.

Figures~\ref{fig: LOUC (Acc) Measure by Normalized Response Time and Positive Predicted Label} and~\ref{fig: LOUC (P) Measure by Normalized Response Time and Positive Predicted Label} present only the decisions labeled (predicted) as matches, and show their distribution by whether it was a true match ($TP$, label with 1) or true non-match ($FP$, labeled with 0). Both figures show that $TP$ decisions are concentrated in the positive LOUC values, whereas $FP$ decisions are more concentrated in the negative LOUC values, which essentially means that correct decisions deteriorate the calibration level, whereas wrong decisions improve it. One possible explanation is that the figure represents the tendency of the annotators towards under-confidence. For correct decisions, accuracy increases and a decrease in calibration can only be explained by confidence levels that are below accuracy (under-confidence). The same is true for wrong decisions, which deteriorate accuracy, and improved calibration levels can be achieved only if confidence levels are below accuracy. Again, the LOUC measure presents, at a higher granularity level, the overall tendency to under-confidence, which was originally presented in Figure~\ref{fig:calibration-acc-by-mean-normalized-confidence}. The clear dichotomy of the measure indicates its predictive powers in distinguishing between correct and wrong decisions.


\section{Related work}
\label{sec: related work}

Schema matching can be formalized as a binary classification task and thus can utilize the classical measures presented here, along with other measures~\cite{do2003comparison}. However, previous works suggested evaluation measures that consider non-binary similarity matrices for schema matching and other data integration tasks \cite{sagi2018non, Sagi20142BON}. 

In~\cite{ackerman2019cognitive, Hung2014PayasyougoRI}, factors affecting human annotators, in contrast to algorithms, were analyzed. Various methods were suggested to construct annotation tasks that direct the annotators into good performance \cite{gal2019learning} and even to improve labeling~\cite{shraga2022powarematch, shraga2022humanal}.

Other works focus on the evaluation of human annotators, including the prediction of matching experts \cite{shraga2021learning} and the investigation of human (meta) cognition-based measures, similar to the calibration measure \cite{hattie2013calibration, hadwin2013calibration}. Additional works test the applicability of human-based measures to algorithms~\cite{ovadia2019can}.
\section{Limitations and Challenges}
\label{sec: Limitations and Challenges}
Schema matching is a core data integration task. Even with multiple algorithmic matchers, human assistance is still required and more research is needed to properly understand human behavior and effectively utilize them in matching. Humans are different from algorithms, which poses challenges and also provides collaborations opportunities for behavioral, computer and data sciences.

The calibration measure, introduced in meta-cognition, helps quantifying the extent to which a matcher can self-assess performance. We present empirical interesting results regarding human-annotated data and present the LOUC measure, a novel calibration-based measure that isolates the effect of each matching question over the overall calibration of an annotator. Escorted by empirical results, we discuss the behavior of the measure and its potential in directing human annotators to improved performance.

LOUC measures the effect of a single decision over the absolute calibration value. Ignoring the calibration sign, we cannot distinguish between under/over confident users. Future work can aim at evaluating the direction of the effect, adding a new dimension to the analysis. Furthermore, the calibration measure uses the mean confidence, a continuous measure, and a traditional evaluation measure, which discretely classifies each decision into groups ($TP$, $TN$, {\em etc}). A promising next step will construct calibration and LOUC measures that utilize continuous (non-binary) evaluation measures.

We also present an initial analysis over schema matching annotated data. Other matching tasks, {\em e.g.}, sentence and entity matching, can take advantage of LOUC and further analysis that tests the consistency of the measure's behavior in different tasks is required. Lastly, we have shown interesting connections between LOUC and other decision attributes, which serve as a basis for developing machine learning models to utilize the measure and its relations with other attributes to predict and improve human-annotated data. 


\bibliographystyle{ACM-Reference-Format}
\bibliography{bibliography.bib}


\begin{thebibliography}{35}


\ifx \showCODEN    \undefined \def \showCODEN     #1{\unskip}     \fi
\ifx \showDOI      \undefined \def \showDOI       #1{#1}\fi
\ifx \showISBNx    \undefined \def \showISBNx     #1{\unskip}     \fi
\ifx \showISBNxiii \undefined \def \showISBNxiii  #1{\unskip}     \fi
\ifx \showISSN     \undefined \def \showISSN      #1{\unskip}     \fi
\ifx \showLCCN     \undefined \def \showLCCN      #1{\unskip}     \fi
\ifx \shownote     \undefined \def \shownote      #1{#1}          \fi
\ifx \showarticletitle \undefined \def \showarticletitle #1{#1}   \fi
\ifx \showURL      \undefined \def \showURL       {\relax}        \fi
\providecommand\bibfield[2]{#2}
\providecommand\bibinfo[2]{#2}
\providecommand\natexlab[1]{#1}
\providecommand\showeprint[2][]{arXiv:#2}

\bibitem[Academic(2023)]%
        {ProlificAcademic}
\bibfield{author}{\bibinfo{person}{Prolific Academic}.}
  \bibinfo{year}{2023}\natexlab{}.
\newblock
\newblock
\urldef\tempurl%
\url{https://www.prolific.co/}
\showURL{%
\tempurl}


\bibitem[Ackerman et~al\mbox{.}(2019)]%
        {ackerman2019cognitive}
\bibfield{author}{\bibinfo{person}{Rakefet Ackerman}, \bibinfo{person}{Avigdor
  Gal}, \bibinfo{person}{Tomer Sagi}, {and} \bibinfo{person}{Roee Shraga}.}
  \bibinfo{year}{2019}\natexlab{}.
\newblock \showarticletitle{A cognitive model of human bias in matching}. In
  \bibinfo{booktitle}{\emph{PRICAI 2019: Trends in Artificial Intelligence:
  16th Pacific Rim International Conference on Artificial Intelligence, Cuvu,
  Yanuca Island, Fiji, August 26--30, 2019, Proceedings, Part I 16}}. Springer,
  \bibinfo{pages}{632--646}.
\newblock


\bibitem[Ackerman et~al\mbox{.}(2016)]%
        {ackerman2016metacognition}
\bibfield{author}{\bibinfo{person}{Rakefet Ackerman}, \bibinfo{person}{Avi
  Parush}, \bibinfo{person}{Fareda Nassar}, {and} \bibinfo{person}{Avraham
  Shtub}.} \bibinfo{year}{2016}\natexlab{}.
\newblock \showarticletitle{Metacognition and system usability: Incorporating
  metacognitive research paradigm into usability testing}.
\newblock \bibinfo{journal}{\emph{Computers in Human Behavior}}
  \bibinfo{volume}{54} (\bibinfo{year}{2016}), \bibinfo{pages}{101--113}.
\newblock


\bibitem[Bellahsene et~al\mbox{.}(2011)]%
        {bellahsene2011evaluating}
\bibfield{author}{\bibinfo{person}{Zohra Bellahsene}, \bibinfo{person}{Angela
  Bonifati}, \bibinfo{person}{Fabien Duchateau}, {and} \bibinfo{person}{Yannis
  Velegrakis}.} \bibinfo{year}{2011}\natexlab{}.
\newblock \bibinfo{booktitle}{\emph{On evaluating schema matching and
  mapping}}.
\newblock \bibinfo{publisher}{Springer}.
\newblock


\bibitem[Bernstein et~al\mbox{.}(2011)]%
        {Bernstein2011}
\bibfield{author}{\bibinfo{person}{Philip~A. Bernstein},
  \bibinfo{person}{Jayant Madhavan}, {and} \bibinfo{person}{Erhard Rahm}.}
  \bibinfo{year}{2011}\natexlab{}.
\newblock \showarticletitle{Generic Schema Matching, Ten Years Later}.
\newblock \bibinfo{journal}{\emph{{PVLDB}}} \bibinfo{volume}{4},
  \bibinfo{number}{11} (\bibinfo{year}{2011}), \bibinfo{pages}{695--701}.
\newblock


\bibitem[Callison-Burch(2009)]%
        {callison2009fast}
\bibfield{author}{\bibinfo{person}{C. Callison-Burch}.}
  \bibinfo{year}{2009}\natexlab{}.
\newblock \showarticletitle{Fast, cheap, and creative: evaluating translation
  quality using Amazon's Mechanical Turk}. In
  \bibinfo{booktitle}{\emph{EMNLP}}.
\newblock


\bibitem[CrowdFlower(2023)]%
        {CrowdFlower}
\bibfield{author}{\bibinfo{person}{CrowdFlower}.}
  \bibinfo{year}{2023}\natexlab{}.
\newblock
\newblock
\urldef\tempurl%
\url{https://agp.iem.technion.ac.il/incognitomatch/SchemaMatching/}
\showURL{%
\tempurl}


\bibitem[Do et~al\mbox{.}(2003)]%
        {do2003comparison}
\bibfield{author}{\bibinfo{person}{Hong-Hai Do}, \bibinfo{person}{Sergey
  Melnik}, {and} \bibinfo{person}{Erhard Rahm}.}
  \bibinfo{year}{2003}\natexlab{}.
\newblock \showarticletitle{Comparison of schema matching evaluations}. In
  \bibinfo{booktitle}{\emph{Web, Web-Services, and Database Systems: NODe 2002
  Web-and Database-Related Workshops Erfurt, Germany, October 7--10, 2002
  Revised Papers 4}}. Springer Berlin Heidelberg, \bibinfo{pages}{221--237}.
\newblock


\bibitem[Do and Rahm(2002)]%
        {do2002coma}
\bibfield{author}{\bibinfo{person}{Hong-Hai Do} {and} \bibinfo{person}{Erhard
  Rahm}.} \bibinfo{year}{2002}\natexlab{}.
\newblock \showarticletitle{COMA—a system for flexible combination of schema
  matching approaches}. In \bibinfo{booktitle}{\emph{VLDB'02: Proceedings of
  the 28th International Conference on Very Large Databases}}. Elsevier,
  \bibinfo{pages}{610--621}.
\newblock


\bibitem[Falconer and Storey(2007)]%
        {Falconer2007}
\bibfield{author}{\bibinfo{person}{Sean~M. Falconer} {and}
  \bibinfo{person}{Margaret{-}Anne~D. Storey}.}
  \bibinfo{year}{2007}\natexlab{}.
\newblock \showarticletitle{A Cognitive Support Framework for Ontology
  Mapping}.
\newblock In \bibinfo{booktitle}{\emph{International Semantic Web Conference,
  {ISWC}}}. \bibinfo{series}{Lecture Notes in Computer Science},
  Vol.~\bibinfo{volume}{4825}. \bibinfo{publisher}{Springer Berlin Heidelberg},
  \bibinfo{pages}{114--127}.
\newblock
\showISBNx{978-3-540-76297-3}


\bibitem[Fan et~al\mbox{.}(2014)]%
        {fan2014hybrid}
\bibfield{author}{\bibinfo{person}{Ju Fan}, \bibinfo{person}{Meiyu Lu},
  \bibinfo{person}{Beng~Chin Ooi}, \bibinfo{person}{Wang-Chiew Tan}, {and}
  \bibinfo{person}{Meihui Zhang}.} \bibinfo{year}{2014}\natexlab{}.
\newblock \showarticletitle{A hybrid machine-crowdsourcing system for matching
  web tables}. In \bibinfo{booktitle}{\emph{2014 IEEE 30th International
  Conference on Data Engineering}}. IEEE, \bibinfo{pages}{976--987}.
\newblock


\bibitem[Gal et~al\mbox{.}(2019)]%
        {gal2019learning}
\bibfield{author}{\bibinfo{person}{Avigdor Gal}, \bibinfo{person}{Haggai
  Roitman}, {and} \bibinfo{person}{Roee Shraga}.}
  \bibinfo{year}{2019}\natexlab{}.
\newblock \showarticletitle{Learning to rerank schema matches}.
\newblock \bibinfo{journal}{\emph{IEEE Transactions on Knowledge and Data
  Engineering}} \bibinfo{volume}{33}, \bibinfo{number}{8}
  (\bibinfo{year}{2019}), \bibinfo{pages}{3104--3116}.
\newblock


\bibitem[Ghazarian and Noorhosseini(2010)]%
        {ghazarian2010automatic}
\bibfield{author}{\bibinfo{person}{Arin Ghazarian} {and}
  \bibinfo{person}{S~Majid Noorhosseini}.} \bibinfo{year}{2010}\natexlab{}.
\newblock \showarticletitle{Automatic detection of users' skill levels using
  high-frequency user interface events}.
\newblock \bibinfo{journal}{\emph{UMUAI}} (\bibinfo{year}{2010}).
\newblock


\bibitem[Guo et~al\mbox{.}(2017)]%
        {guo2017calibration}
\bibfield{author}{\bibinfo{person}{Chuan Guo}, \bibinfo{person}{Geoff Pleiss},
  \bibinfo{person}{Yu Sun}, {and} \bibinfo{person}{Kilian~Q Weinberger}.}
  \bibinfo{year}{2017}\natexlab{}.
\newblock \showarticletitle{On calibration of modern neural networks}. In
  \bibinfo{booktitle}{\emph{International conference on machine learning}}.
  PMLR, \bibinfo{pages}{1321--1330}.
\newblock


\bibitem[Hadwin and Webster(2013)]%
        {hadwin2013calibration}
\bibfield{author}{\bibinfo{person}{Allyson~F Hadwin} {and}
  \bibinfo{person}{Elizabeth~A Webster}.} \bibinfo{year}{2013}\natexlab{}.
\newblock \showarticletitle{Calibration in goal setting: Examining the nature
  of judgments of confidence}.
\newblock \bibinfo{journal}{\emph{Learning and Instruction}}
  \bibinfo{volume}{24} (\bibinfo{year}{2013}), \bibinfo{pages}{37--47}.
\newblock


\bibitem[Hattie(2013)]%
        {hattie2013calibration}
\bibfield{author}{\bibinfo{person}{John Hattie}.}
  \bibinfo{year}{2013}\natexlab{}.
\newblock \showarticletitle{Calibration and confidence: Where to next?}
\newblock \bibinfo{journal}{\emph{Learning and instruction}}
  \bibinfo{volume}{24} (\bibinfo{year}{2013}), \bibinfo{pages}{62--66}.
\newblock


\bibitem[Hung et~al\mbox{.}(2014)]%
        {Hung2014PayasyougoRI}
\bibfield{author}{\bibinfo{person}{Nguyen Quoc~Viet Hung},
  \bibinfo{person}{Thanh~Tam Nguyen}, \bibinfo{person}{Zolt{\'a}n Mikl{\'o}s},
  \bibinfo{person}{Karl Aberer}, \bibinfo{person}{Avigdor Gal}, {and}
  \bibinfo{person}{Matthias Weidlich}.} \bibinfo{year}{2014}\natexlab{}.
\newblock \showarticletitle{Pay-as-you-go reconciliation in schema matching
  networks}.
\newblock \bibinfo{journal}{\emph{2014 IEEE 30th International Conference on
  Data Engineering}} (\bibinfo{year}{2014}), \bibinfo{pages}{220--231}.
\newblock


\bibitem[InCognitoMatch(2023)]%
        {inCognitoMatch}
\bibfield{author}{\bibinfo{person}{InCognitoMatch}.}
  \bibinfo{year}{2023}\natexlab{}.
\newblock
\newblock
\urldef\tempurl%
\url{https://agp.iem.technion.ac.il/incognitomatch/SchemaMatching/}
\showURL{%
\tempurl}


\bibitem[Koutras et~al\mbox{.}(2021)]%
        {koutras2021valentine}
\bibfield{author}{\bibinfo{person}{Christos Koutras}, \bibinfo{person}{George
  Siachamis}, \bibinfo{person}{Andra Ionescu}, \bibinfo{person}{Kyriakos
  Psarakis}, \bibinfo{person}{Jerry Brons}, \bibinfo{person}{Marios
  Fragkoulis}, \bibinfo{person}{Christoph Lofi}, \bibinfo{person}{Angela
  Bonifati}, {and} \bibinfo{person}{Asterios Katsifodimos}.}
  \bibinfo{year}{2021}\natexlab{}.
\newblock \showarticletitle{Valentine: Evaluating matching techniques for
  dataset discovery}. In \bibinfo{booktitle}{\emph{2021 IEEE 37th International
  Conference on Data Engineering (ICDE)}}. IEEE, \bibinfo{pages}{468--479}.
\newblock


\bibitem[Li(2017)]%
        {li2017human}
\bibfield{author}{\bibinfo{person}{Guoliang Li}.}
  \bibinfo{year}{2017}\natexlab{}.
\newblock \showarticletitle{Human-in-the-loop data integration}.
\newblock \bibinfo{journal}{\emph{Proceedings of the VLDB Endowment}}
  \bibinfo{volume}{10}, \bibinfo{number}{12} (\bibinfo{year}{2017}),
  \bibinfo{pages}{2006--2017}.
\newblock


\bibitem[Li et~al\mbox{.}(2019)]%
        {li2019user}
\bibfield{author}{\bibinfo{person}{Huanyu Li}, \bibinfo{person}{Zlatan
  Dragisic}, \bibinfo{person}{Daniel Faria}, \bibinfo{person}{Valentina
  Ivanova}, \bibinfo{person}{Ernesto Jim{\'e}nez-Ruiz},
  \bibinfo{person}{Patrick Lambrix}, {and} \bibinfo{person}{Catia Pesquita}.}
  \bibinfo{year}{2019}\natexlab{}.
\newblock \showarticletitle{User validation in ontology alignment: functional
  assessment and impact}.
\newblock \bibinfo{journal}{\emph{The Knowledge Engineering Review}}
  \bibinfo{volume}{34} (\bibinfo{year}{2019}).
\newblock


\bibitem[McCann et~al\mbox{.}(2008)]%
        {McCann2008}
\bibfield{author}{\bibinfo{person}{Robert McCann}, \bibinfo{person}{Warren
  Shen}, {and} \bibinfo{person}{AnHai Doan}.} \bibinfo{year}{2008}\natexlab{}.
\newblock \showarticletitle{Matching schemas in online communities: A web 2.0
  approach}. In \bibinfo{booktitle}{\emph{2008 IEEE 24th international
  conference on data engineering}}. IEEE, \bibinfo{pages}{110--119}.
\newblock


\bibitem[OntoBuilder(2023)]%
        {ontobuilder}
\bibfield{author}{\bibinfo{person}{OntoBuilder}.}
  \bibinfo{year}{2023}\natexlab{}.
\newblock
\newblock
\urldef\tempurl%
\url{https://github.com/shraga89/Ontobuilder-Research-Environment}
\showURL{%
\tempurl}


\bibitem[Ovadia et~al\mbox{.}(2019)]%
        {ovadia2019can}
\bibfield{author}{\bibinfo{person}{Yaniv Ovadia}, \bibinfo{person}{Emily
  Fertig}, \bibinfo{person}{Jie Ren}, \bibinfo{person}{Zachary Nado},
  \bibinfo{person}{David Sculley}, \bibinfo{person}{Sebastian Nowozin},
  \bibinfo{person}{Joshua Dillon}, \bibinfo{person}{Balaji Lakshminarayanan},
  {and} \bibinfo{person}{Jasper Snoek}.} \bibinfo{year}{2019}\natexlab{}.
\newblock \showarticletitle{Can you trust your model's uncertainty? evaluating
  predictive uncertainty under dataset shift}.
\newblock \bibinfo{journal}{\emph{Advances in neural information processing
  systems}}  \bibinfo{volume}{32} (\bibinfo{year}{2019}).
\newblock


\bibitem[Ross et~al\mbox{.}(2010)]%
        {Ross2010}
\bibfield{author}{\bibinfo{person}{Joel Ross}, \bibinfo{person}{Lilly Irani},
  \bibinfo{person}{M Silberman}, \bibinfo{person}{Andrew Zaldivar}, {and}
  \bibinfo{person}{Bill Tomlinson}.} \bibinfo{year}{2010}\natexlab{}.
\newblock \showarticletitle{Who are the crowdworkers?: shifting demographics in
  mechanical turk}. In \bibinfo{booktitle}{\emph{CHI'10 Extended Abstracts on
  Human Factors in Computing Systems}}. ACM, \bibinfo{pages}{2863--2872}.
\newblock


\bibitem[Sagi(2014)]%
        {Sagi20142BON}
\bibfield{author}{\bibinfo{person}{Tomer Sagi}.}
  \bibinfo{year}{2014}\natexlab{}.
\newblock \showarticletitle{2B or not 2B and everything in between — novel
  evaluation methods for matching problems}.
\newblock \bibinfo{journal}{\emph{2014 IEEE 30th International Conference on
  Data Engineering Workshops}} (\bibinfo{year}{2014}),
  \bibinfo{pages}{325--329}.
\newblock


\bibitem[Sagi and Gal(2018)]%
        {sagi2018non}
\bibfield{author}{\bibinfo{person}{Tomer Sagi} {and} \bibinfo{person}{Avigdor
  Gal}.} \bibinfo{year}{2018}\natexlab{}.
\newblock \showarticletitle{Non-binary evaluation measures for big data
  integration}.
\newblock \bibinfo{journal}{\emph{The VLDB Journal}} \bibinfo{volume}{27},
  \bibinfo{number}{1} (\bibinfo{year}{2018}), \bibinfo{pages}{105--126}.
\newblock


\bibitem[Sarasua et~al\mbox{.}(2012)]%
        {Crowdmap}
\bibfield{author}{\bibinfo{person}{C. Sarasua}, \bibinfo{person}{E. Simperl},
  {and} \bibinfo{person}{N.~F Noy}.} \bibinfo{year}{2012}\natexlab{}.
\newblock \showarticletitle{Crowdmap: Crowdsourcing ontology alignment with
  microtasks}. In \bibinfo{booktitle}{\emph{ISWC}}.
\newblock


\bibitem[Shraga(2020)]%
        {shraga2020mind}
\bibfield{author}{\bibinfo{person}{Roee Shraga}.}
  \bibinfo{year}{2020}\natexlab{}.
\newblock \showarticletitle{Mind over Matter: Humans In and Humans Out in
  Matching}. In \bibinfo{booktitle}{\emph{Proceedings of the VLDB 2020 PhD
  Workshop co-located with the 46th International Conference on Very Large
  Databases (VLDB 2020), ONLINE}}.
\newblock


\bibitem[Shraga(2022)]%
        {shraga2022humanal}
\bibfield{author}{\bibinfo{person}{Roee Shraga}.}
  \bibinfo{year}{2022}\natexlab{}.
\newblock \showarticletitle{HumanAL: calibrating human matching beyond a single
  task}. In \bibinfo{booktitle}{\emph{Proceedings of the Workshop on
  Human-In-the-Loop Data Analytics}}. \bibinfo{pages}{1--8}.
\newblock


\bibitem[Shraga et~al\mbox{.}(2021)]%
        {shraga2021learning}
\bibfield{author}{\bibinfo{person}{Roee Shraga}, \bibinfo{person}{Ofra Amir},
  {and} \bibinfo{person}{Avigdor Gal}.} \bibinfo{year}{2021}\natexlab{}.
\newblock \showarticletitle{Learning to Characterize Matching Experts}. In
  \bibinfo{booktitle}{\emph{2021 IEEE 37th International Conference on Data
  Engineering (ICDE)}}. IEEE, \bibinfo{pages}{1236--1247}.
\newblock


\bibitem[Shraga and Gal(2022)]%
        {shraga2022powarematch}
\bibfield{author}{\bibinfo{person}{Roee Shraga} {and} \bibinfo{person}{Avigdor
  Gal}.} \bibinfo{year}{2022}\natexlab{}.
\newblock \showarticletitle{PoWareMatch: a Quality-aware Deep Learning Approach
  to Improve Human Schema Matching}.
\newblock \bibinfo{journal}{\emph{ACM Journal of Data and Information Quality
  (JDIQ)}} \bibinfo{volume}{14}, \bibinfo{number}{3} (\bibinfo{year}{2022}),
  \bibinfo{pages}{1--27}.
\newblock


\bibitem[Shraga et~al\mbox{.}(2020a)]%
        {shraga2020adnev}
\bibfield{author}{\bibinfo{person}{Roee Shraga}, \bibinfo{person}{Avigdor Gal},
  {and} \bibinfo{person}{Haggai Roitman}.} \bibinfo{year}{2020}\natexlab{a}.
\newblock \showarticletitle{Adnev: Cross-domain schema matching using deep
  similarity matrix adjustment and evaluation}.
\newblock \bibinfo{journal}{\emph{Proceedings of the VLDB Endowment}}
  \bibinfo{volume}{13}, \bibinfo{number}{9} (\bibinfo{year}{2020}),
  \bibinfo{pages}{1401--1415}.
\newblock


\bibitem[Shraga et~al\mbox{.}(2020b)]%
        {shraga2020incognitomatch}
\bibfield{author}{\bibinfo{person}{Roee Shraga}, \bibinfo{person}{Coral
  Scharf}, \bibinfo{person}{Rakefet Ackerman}, {and} \bibinfo{person}{Avigdor
  Gal}.} \bibinfo{year}{2020}\natexlab{b}.
\newblock \showarticletitle{Incognitomatch: Cognitive-aware matching via
  crowdsourcing}. In \bibinfo{booktitle}{\emph{Proceedings of the 2020 ACM
  SIGMOD international conference on management of data}}.
  \bibinfo{pages}{2753--2756}.
\newblock


\bibitem[Zhang et~al\mbox{.}(2018)]%
        {zhang2018reducing}
\bibfield{author}{\bibinfo{person}{Chen Zhang}, \bibinfo{person}{Lei Chen},
  \bibinfo{person}{HV Jagadish}, \bibinfo{person}{Mengchen Zhang}, {and}
  \bibinfo{person}{Yongxin Tong}.} \bibinfo{year}{2018}\natexlab{}.
\newblock \showarticletitle{Reducing Uncertainty of Schema Matching via
  Crowdsourcing with Accuracy Rates}.
\newblock \bibinfo{journal}{\emph{IEEE Transactions on Knowledge and Data
  Engineering}} (\bibinfo{year}{2018}).
\newblock


\end{thebibliography}

\end{document}